\documentclass[aps,pre,preprint,showkeys,a4paper]{revtex4-1}
\usepackage[utf8]{inputenc}
\pdfoutput=1
\usepackage{booktabs}
\usepackage{graphicx,xcolor}

\begin{document}

\title{Classification of particle trajectories in living cells: machine learning versus statistical testing hypothesis for fractional anomalous diffusion}
\author{Joanna Janczura}
\author{Patrycja Kowalek}
\author{Hanna Loch-Olszewska}
\author{Janusz Szwabiński}
\author{Aleksander Weron}
\affiliation{Faculty of Pure and Applied Mathematics, Hugo Steinhaus Center, Wrocław University of Science and Technology, 50-370 Wrocław, Poland}

\begin{abstract}
Single-particle tracking (SPT) has become a popular tool to study the intracellular transport of molecules in living cells. Inferring the character of their dynamics  is important, because it determines the organization and functions of the cells. For this reason, one of the first steps in the analysis of SPT data is the identification of the diffusion type of the observed particles. The most popular method to identify the class of a trajectory is based on the mean square displacement (MSD). However, due to its known limitations, several other approaches have been already proposed. With the recent advances in algorithms and the developments of modern hardware, the classification attempts rooted in machine learning (ML) are of particular interest. In this work, we adopt two ML ensemble algorithms, i.e. random forest and gradient boosting, to the problem of trajectory classification. We present a new set of features used to transform the raw trajectories data into input vectors required by the classifiers. The resulting models are then applied to real data for G protein-coupled receptors and G proteins. The classification results are compared to recent statistical methods going beyond MSD.
\end{abstract}


\maketitle

\section{Introduction}

Single-particle tracking (SPT) has become an important tool in the biophysical community in recent years. It was first carried out on proteins diffusing in the cell membrane~\cite{BAR82,KUS93}. Since then it was successfully used to study different transport processes in intracellular environment, providing valuable information about mechano-structural characteristics of living cells. For instance, it helped already to unveil the details of the movement of molecular motors inside cells~\cite{YIL03,KUR05} or of target search mechanisms of nuclear proteins~\cite{IZE14}.

Living cells belong to the class of active systems~\cite{GAL13}, in which the particles undergo simultaneous active and thermally driven transport. It has been shown already  that the dynamics of proteins in cells determines their organization and functions~\cite{BRE14}. This is the reason why it is crucial to identify the type of motion of the observed particles in order to deduct their driving forces~\cite{SAX94,SAX97,FED96,MET00}.

Over the last decades, a number of stochastic models has been already proposed to describe the intracellular transport of molecules~\cite{MIC10,MET00}. Within those models, the dynamics of molecules usually alternates between distinct types of diffusion, each of which may be associated with a different physical scenario. The Brownian motion~\cite{ALV16} models a particle that diffuses freely, i.e. it does not meet any obstacles in its path nor it interacts with other molecules in its surrounding. The subdiffusion is appropriate to represent trapped particles~\cite{HOZ12,MET00}, particles which encounter fixed or moving obstacles~\cite{SAX94,BER14} or particles slowed down due to the viscoelastic properties of the cytoplasm~\cite{WEI04}. Finally, the superdiffusion models the motion driven by molecular motors: the particles move faster than in a free diffusion case and in a specific direction~\cite{ARC08}. The sub- and superdiffusion together are often referred to as the anomalous diffusion.

The standard method of classification of individual trajectories into those three types of diffusion is based on the mean square displacement (MSD)~\cite{MIC10}. Within this approach one fits the theoretical MSD curves for various models to the data and then selects the best fit with statistical analysis~\cite{MON12}. A linear MSD curve indicates the free diffusion, a sublinear (superlinear) one - the subdiffusion (the superdiffusion). However, there are some issues related with this method. In many cases, the experimental trajectories are too short to extract a meaningful information from MSD. Moreover, the finite precision adds a term to the MSD, which is known to limit the interpretation of the data~\cite{MIC10,SAX97,KEP15,BRI18}. As a result, several methods improving or going beyond the MSD have been introduced to overcome these problems. For instance, Michalet~\cite{MIC10} used an iterative method called the optimal least-squares fit to determine the optimal number of points to obtain the best fit to MSD in the presence of localization errors. Weiss~\cite{WEI19} used a resampling approach that eliminates localization errors in the time-averaged MSD of subdiffusive fractional Brownian motion processes. The trajectory spread in space calculated through the radius of gyration~\cite{SAX93}, the Van Hove displacements distributions and deviations from Gaussian statistics~\cite{VAL01}, self-similarity of trajectory using different powers of the displacement~\cite{GAL10}, velocity autocorrelation function~\cite{GRE13,FUL17} or the time-dependent directional persistence of trajectories~\cite{RAU07} methods can be combined with the output of MSD to improve the classification results. The distribution of directional changes~\cite{BUR13}, the mean maximum excursion method~\cite{TEJ10} and the fractionally integrated moving average (FIMA) framework~\cite{BUR15} may efficiently replace the MSD estimator for classification purposes. Hidden Markov Models (HMM) has been proposed to check the heterogeneity within single trajectories~\cite{DAS09,SLA15}. They have  proven to be quite useful in the detection of confinement~\cite{SLA18}. Last but not least, classification based on hypothesis testing, both relying on MSD and going beyond this statistics, has been shown to be quite successful as well~\cite{BRI18,WER19}.

An alternative, very promising approach to SPT data analysis is rooted in computer science. Namely, classification of trajectories may be seen as a subject of machine learning (ML)~\cite{RAS15}. In the ML context, classification relies on available data, because its goal is to identify to which category a new observation belongs on the basis of a training data set containing observations with a known category membership.  

There is already a number of attempts to analyze particle trajectories with machine learning methods. Among them, Bayesian approach~\cite{MON12,THA18,CHE19}, random forests~\cite{WAG17,KOW19,MUN20}, neural networks~\cite{DOS16} and deep neural networks~\cite{KOW19,GRA19,BO19,HAN20} have gained a lot of attention and popularity. While some of the works have focused just on the identification of the diffusion modes~\cite{DOS16,WAG17,KOW19}, the others went beyond just the classification of diffusion and tried to extract quantitative information about the trajectories (e.g. the anomalous exponent~\cite{MUN20,GRA19}).

Recently, we presented a comparison of performance of two different classes of methods: traditional feature-based algorithms (random forest and gradient boosting) and a modern deep learning approach based on convolutional neural networks~\cite{KOW19}. The latter constitutes nowadays the state-of-the-art technology for automatic data classification and is much simpler to use from the perspective of the end-user, because it operates on raw data and does not require any preprocessing effort from human experts~\cite{HAT18}. In contrast, the traditional methods require a representation of trajectories by a set of human-engineered features or attributes~\cite{MIT97}. In most of the applications the deep learning approach outperforms the traditional methods. However, in some situations it is still worth to use them, because they usually work better on small data sets, are computationally cheaper and easier to interpret. From our results it follows that both approaches achieve excellent (and very similar) accuracies on synthetic data. But they turned out to be really bad in terms of transfer learning. This concept refers to a situation, in which a classifier is trained in one setting and then applied to a different one. The classifiers from Ref.~\cite{KOW19} were not able to successfully classify trajectories generated with methods different from the ones used for the training set. 

In this paper, we are going to present an improved version of the traditional classifiers presented in Ref.~\cite{KOW19}. We will propose a new set of training data as well as a new collection of features describing a trajectory. Both are inspired by a recent statistical analysis of anomalous diffusion~\cite{WER19}. To illustrate the transfer learning abilities of the new classifiers, we will apply them to the data from a single-particle tracking experiment on G protein-coupled receptors and G proteins~\cite{SUN17}. Results of classification from Ref.~\cite{WER19} will be used as a benchmark.

The paper is organized as follows. In Sec.~\ref{sec:diff}, we briefly introduce the different modes of diffusion and methods of their analysis. Sec.~\ref{sec:ml} contains a short description of the machine learning methods used in this work. Stochastic models of diffusion for generation of synthetic data are presented in Sec.~\ref{sec:stochastic}. The data itself is characterized in Sec.~\ref{sec:data}. The set of features used as input to the classifiers is introduced in Sec.~\ref{sec:feature}. Our results are presented in Sec.~\ref{sec:results}, followed by some concluding remarks.

\section{Diffusion modes and their analysis}
\label{sec:diff}

As already mentioned in the introduction, identification of the diffusion modes of particles within living cells is important, because they reflect the interactions of those particles with their surrounding. For instance, if a particle is driven by a free diffusion (Brownian motion)~\cite{ALV16}, we expect that it does not meet any obstacles in its path and does not undergo any relevant interactions with other particles. Deviations from Brownian motion are called anomalous diffusion and can be divided into two distinct classes. Subdiffusion is slower than the normal one. It usually occurs in crowded or constrained domains and can be brought together with different physical mechanisms including immobile obstacles, cytoplasm viscosity, crowding, trapping and heterogeneities~\cite{HOF13,SZY09,JEO10}. Superdiffusion represents active transport along the cytoskeleton, assisted by molecular motors~\cite{ARC08}. Particles undergoing that type of motion move faster than those freely diffusing and usually do not come back to previous positions.

Although different scenarios for both classes of anomalous diffusion are possible~\cite{MET00,WEI11,WEI12,HEL11,SZY09,SAD17}, for the purpose of this work we will limit ourselves to those three basic types mentioned above: free, sub- and superdiffusion.

The most popular method of deducing a particles' type of motion from their trajectories is based on the analysis of the mean square displacement (MSD)~\cite{QIA91},
\begin{equation}
MSD(t) = \mathrm{E}\left( \Vert X_{t+t_0} -X_{t_0}\Vert^2 \right),
\label{eq:msd}
\end{equation}
where $(X_t)_{t>0}$ is a particle trajectory, $\Vert \cdot \Vert$ is the Euclidean norm and $\mathrm{E}$ is the expectation of the probability space. Since 
 in many experiments only a limited number of trajectories is observed, the time averaged MSD (TAMSD) calculated from a single trajectory is usually used as the estimator of MSD,
\begin{equation}
\widehat{MSD}(n\Delta t) =\frac{1}{N-n+1}\sum_{i=0}^{N-n} \Vert X_{t_{i+n}}-X_{t_i} \Vert^2.
\label{eq:tamsd}
\end{equation} 
The trajectory is assumed to be given in form of $N$ consecutive two dimensional positions $X_i =(x_i,y_i)$ ($i=0,\dots,N$) recorded with a constant time interval $\Delta t$ and $n$ is the time lag between the initial and the final positions of the particle.  If the underlying process is ergodic and has stationary increments, TAMSD converges to the theoretical MSD~\cite{WEI11}.

TAMSD as a function of the time lag for the normal diffusion converges asymptotically to a linear function~\cite{SAX97}, i.e. for large $N$:
\begin{equation}
\widehat{MSD}(n \Delta t) \sim  4D(n\Delta t),
\label{eq:mmsd}
\end{equation}
with $D$ being the diffusion coefficient. For subdiffusion, being slower than diffusion, the behaviour of TAMSD is sublinear, while for superdiffusion, being faster than diffusion, the behaviour is superlinear. 
Thus, for pure trajectories with no localization errors one could easily determine their diffusion type by fitting a function $\alpha \log (n\Delta t) +\beta$ to the estimated $\log [\widehat{MSD}(n\Delta t)]$ curve. If $\alpha<1$ then the trajectory can be identified as subdiffusive, while if $\alpha>1$, as superdiffusive. Although theoretically this approach allows for the uncomplicated distinction of the diffusion types, there are several issues related with it as a method for classification.  First, real trajectories are usually noisy, which makes the fitting of a mathematical model a challenging task, even in the simplest case of the normal diffusion~\cite{MIC10,WEI19}. Secondly, according to Eq.~(\ref{eq:tamsd}), only the values of $\widehat{MSD}$ corresponding to small time lags are well averaged. The larger the lag, the smaller is the number of displacements contributing to the averages, resulting in fluctuations increasing with the lag. Selecting a suitable lag is by the way a well known problem in biophysics~\cite{BRI18,VEG18,LAN18a}. 
Since many real trajectories are short, we are forced to concentrated on short times (small lags). This induces another problem in a classification method based only on MSD curves, as in this case the different power laws look alike even in the absence of noise.

\section{Machine learning approach}
\label{sec:ml}

Several different procedures have been already proposed to circumvent the limitations of the MSD~\cite{MIC10,SAX93,VAL01,GAL10,RAU07,BUR13,TEJ10,DAS09,SLA15,SLA18,BRI18,WER19}, including the use of machine learning methods~\cite{MON12,THA18,CHE19,WAG17,KOW19,MUN20,GRA19,BO19}. Recently, we discussed the applicability of three different machine learning algorithms to classification, including two feature-based methods and a deep learning one~\cite{KOW19}. The results of that study were ambiguous. On one hand all of the methods performed excellent on the test data, on the other - they failed to transfer their knowledge to data coming from unseen physical models. The latter finding practically disqualified them as candidates for a reliable classification tool.

In this paper, we are going to continue the analysis started in Ref.~\cite{KOW19} and present improved versions of the classifiers, which performs much better in terms of transfer learning. We will focus on the traditional machine learning methods: the random forest (RF)~\cite{HO95,HO98} and the gradient boosting (GB)~\cite{SCH98,FRI02}. Both methods are feature-based, meaning that each instance in the data set is described by a set of human-engineered attributes~\cite{MIT97}. And both belong to the class of ensemble methods, which combine multiple base classifiers to form a better one. In each case, decision trees~\cite{SON15} are used as the base classifiers. 

A decision tree is built by splitting the original dataset (trajectories with known classes), constituting the root node of the tree, into subsets, which represent the successor children. The splitting is based on a set of rules utilizing the values of features. This process is repeated on each derived subset in a recursive manner. The recursion is completed when the subset at a node has all samples belonging to the same class (i.e. the node is pure) or when splitting no longer adds value to the classification. At each step, a feature that best splits the data is chosen. Two metrics are typically used  to measure the quality of the split: Gini impurity and information gain~\cite{RAS15}. 

Gini impurity tells us how often a randomly chosen element from the set would be incorrectly labeled if it was randomly labeled according to the distribution of labels in that set. It is given by
\begin{equation}
I_G = \sum_{i=1}^J p_i (1-p_i),
\label{eq:gini}    
\end{equation}
where $J$ is the number of classes ($J=3$ in our case) and $p_i$ is the fraction of items labeled with class $i$ in the set.

Information gain related to a split is simply the reduction of information entropy~\cite{SHA48}, calculated as the difference between the entropy of a parent node in the tree and a weighted sum of entropies of its children nodes. The entropy itself is given as
\begin{equation}
    H=-\sum_{i=1}^J p_i \log_2 p_i,
\end{equation}
where $p_1,~p_2,\dots$ are fractions of each class present in the node.

Decision trees are often used for classification purposes, because they are easy to understand and interpret. However, single trees are unstable in the sense that a small variation in the data may lead to a completely different tree~\cite{GAR13}. They also have a tendency to overfit, i.e. they model the training data too well and learn  noise or random fluctuations as meaningful concepts, which limits their accuracy in case of unseen data~\cite{BRA13}. That is why they are rather used as building blocks of the ensembles and not as stand alone classifiers. 

In a random forest, multiple decision trees are constructed independently from the same training data. The predictions of individual trees are aggregated and then their mode is taken as the final output. In gradient boosting, the trees are not independent. Instead, they are built sequentially by learning from mistakes commited by the ensemble. In many applications, gradient boosting is expected to have a better performance than random forest. However, it is usually not the better choice in case of very noisy data. 

\begin{figure}
\includegraphics[scale=0.5]{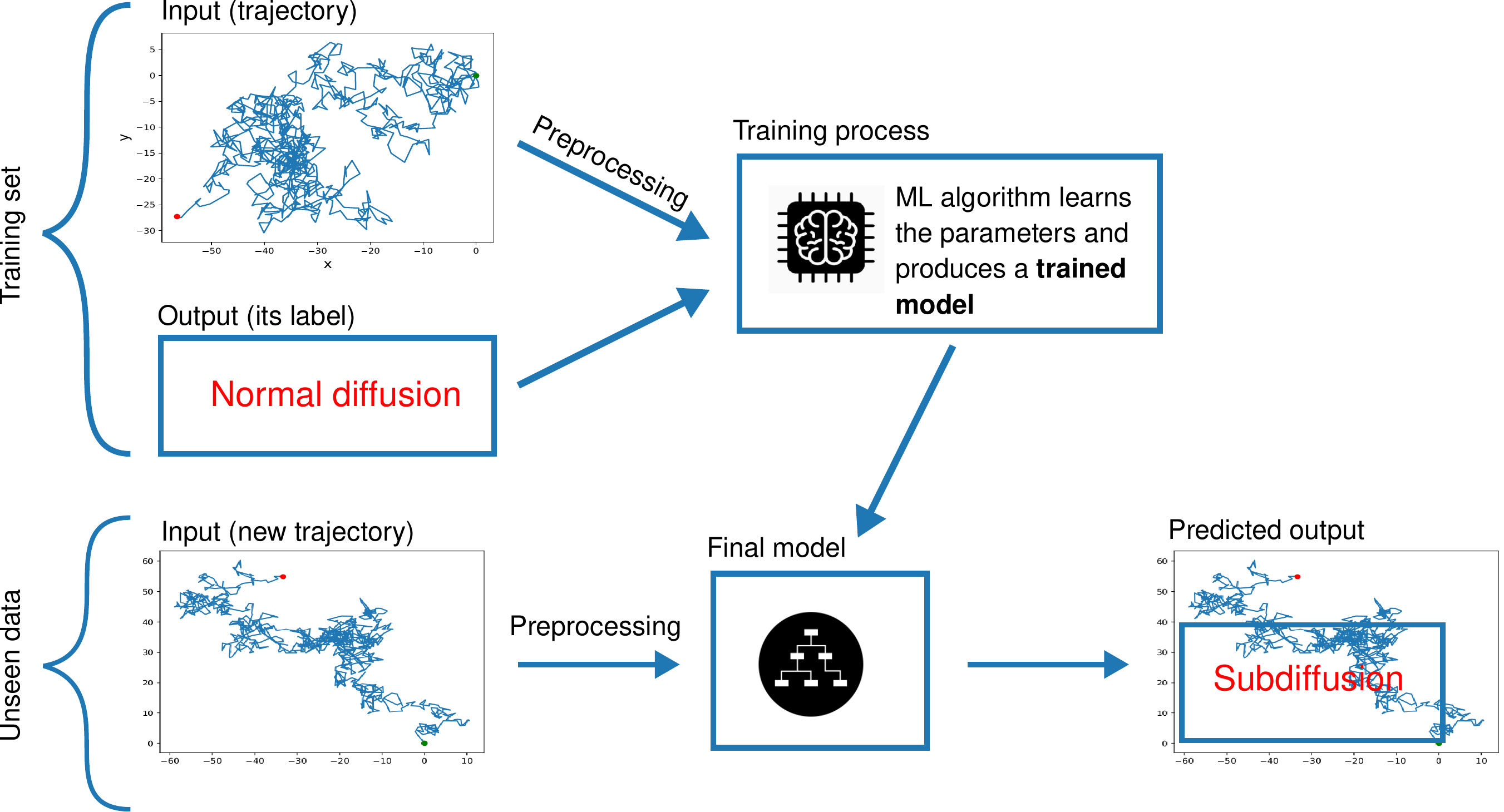}
\caption{Workflow of our classification method. The training set is composed of a large number of synthetic trajectories (Sec.~\ref{sec:synthetic}). The preprocessing phase consists in extraction of features introduced in Sec.~\ref{sec:feature}. \label{fig:workflow}}
\end{figure}

A workflow of our classification method is shown in Fig.~\ref{fig:workflow}. The training set consists of a large number of synthetic trajectories and their labels (diffusion modes). The trajectories were generated with various kinds of theoretical models of diffusion (see Sections~\ref{sec:stochastic} and~\ref{sec:synthetic} for further details).
In the preprocessing phase, the raw data is cleaned and transformed into a form required as input by the classifier. Many traditional classifiers including random forest and gradient boosting work much better with vectors of features characterizing each trajectory instead of raw data. The features used in this work are introduced in Sec.~\ref{sec:feature}. Some authors normalize the trajectories before further processing~\cite{MUN20}. However, we omitted this step as our preliminary analysis indicated a significant decrease in the performance of the classifiers induced by normalization. The ensembles of trees were inferred from the feature vectors and their labels. Once trained, they may be used to classify new trajectories, including the experimental ones.

\section{Stochastic models of diffusion}
\label{sec:stochastic}

The most popular theoretical models of diffusion commonly employed are: continuous-time random walk (CTRW)~\cite{MET00}, obstructed diffusion (OD)~\cite{HAV87,SAX94}, random walk on random walks (RWRW)~\cite{BLO19}, random walks on percolating clusters (RWPC)~\cite{STR80,MET14} fractional Brownian motion (FBM)~\cite{MAN68,GUI07,BUR12}, fractional Levy $\alpha$-stable motion (FLSM)~\cite{BUR10}, fractional Langevin equation (FLE)~\cite{KOU04} and autoregressive fractionally integrated moving average (ARFIMA)~\cite{BUR14}. They are applicable to different physical environments: trapping and crowded environments (CTRW, FFPE); labyrinthine environments (OD, RWPC, RWRW); viscoelastic systems (FBM, FLSM, FLE, ARFIMA); systems with time-dependent diffusion (scaled FBM, ARFIMA).
Following Refs.~\cite{BRI18,WER19}, we will focus on three stochastic processes known to generate different kinds of fractional diffusion: fractional Brownian motion, directed Brownian motion (DBM)~\cite{ELS00} and Ornstein-Uhlenbeck process (OU)~\cite{MAC10}.

\begin{figure}
\includegraphics[scale=0.5]{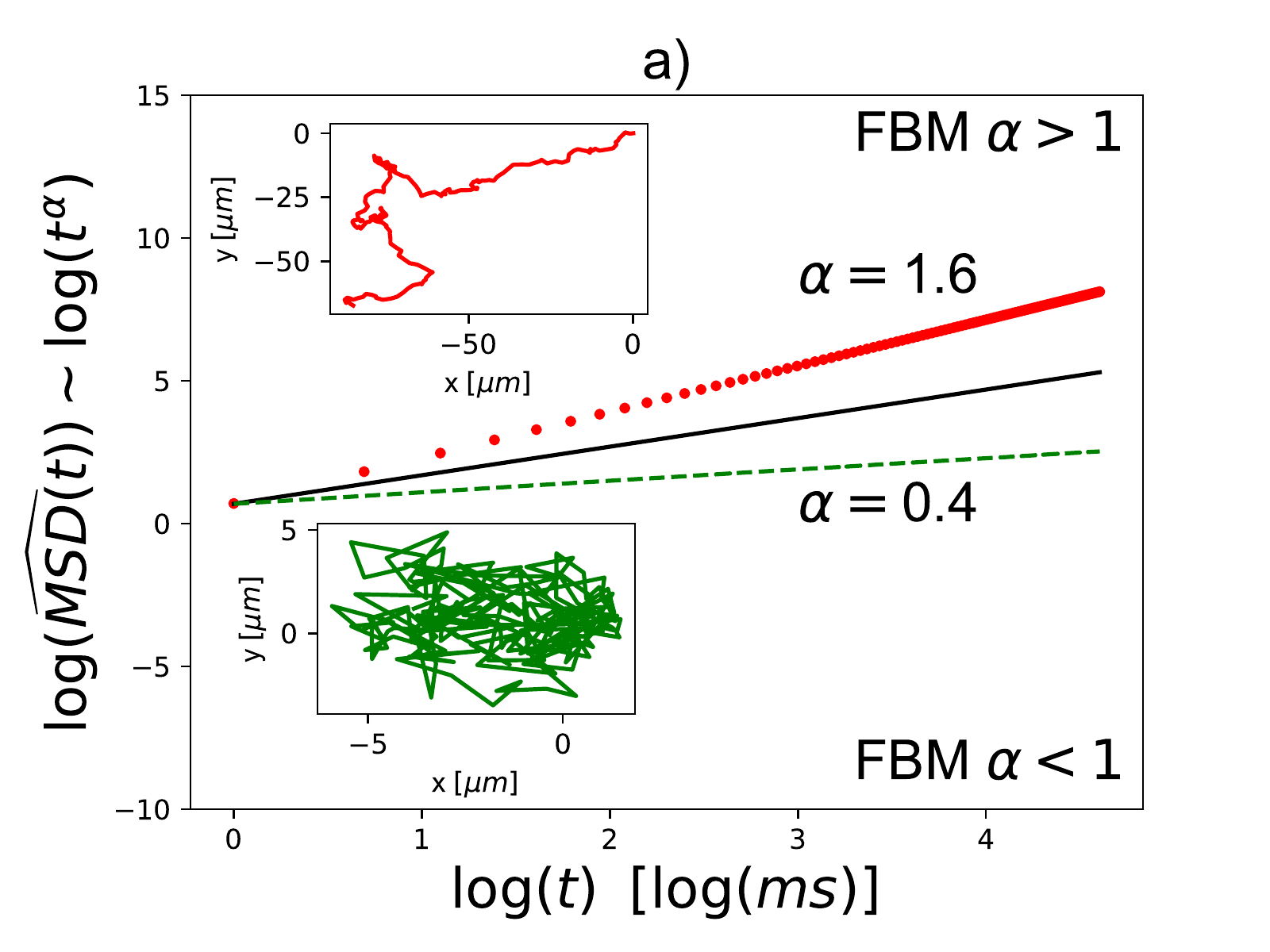}
\includegraphics[scale=0.5]{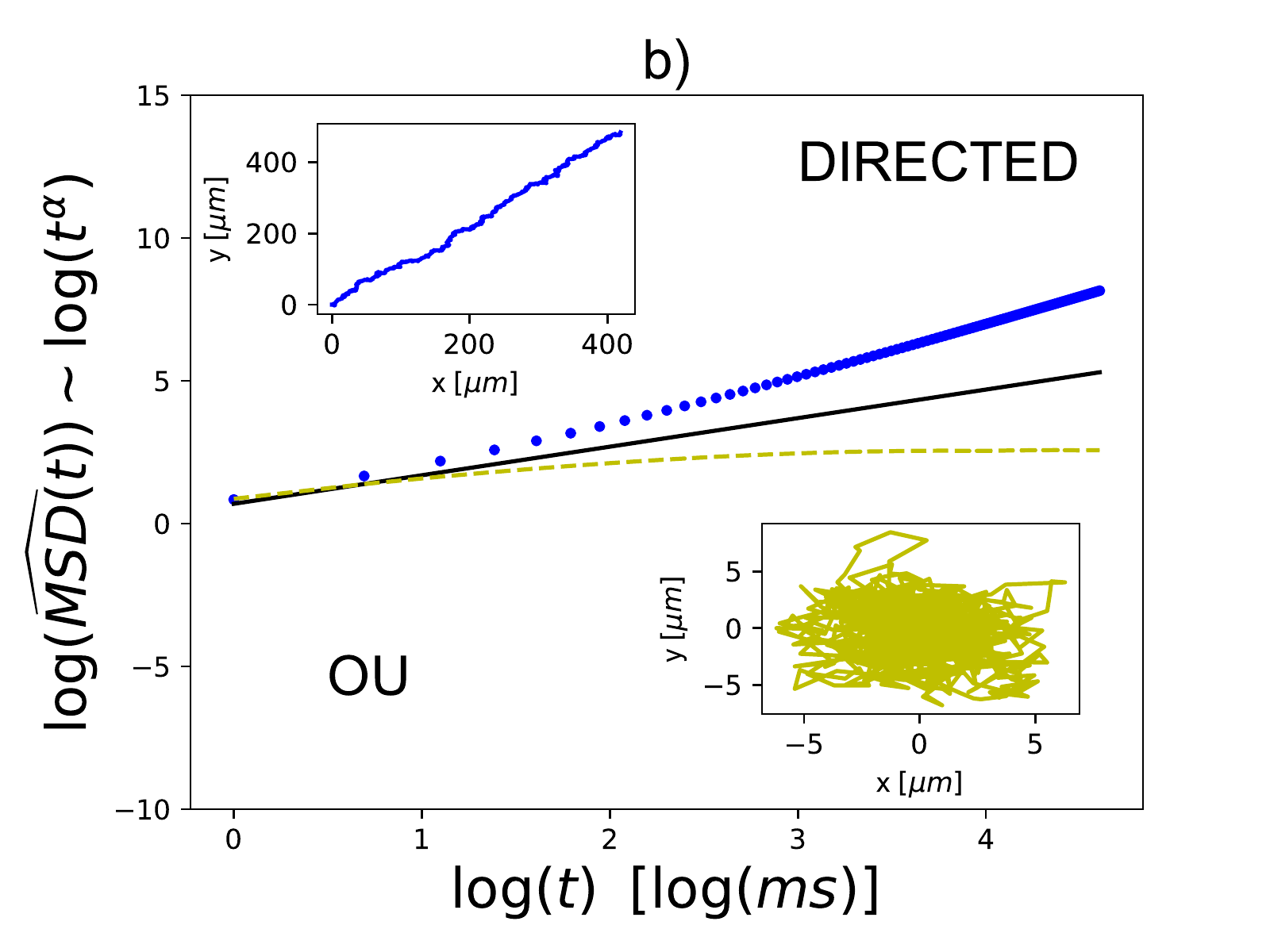}
\caption{Time-averaged mean-squared displacement calculated for: (a) FBM with different values of $\alpha$, (b) DBM and OU processes. The trajectories used to calculate the MSD curves are shown in the corresponding insets and are consistent with real data time and distance scales: $ms$ and $\mu m$ accordingly. The solid line in both plots indicates TAMSD of normal diffusion.  \label{fig:example_traj}}
\end{figure}

FBM is the solution of the stochastic differential equation
\begin{equation}
dX_t^i = \sigma dB_t^{H,i},~~i=1,2,
\end{equation}
where the parameter $\sigma>0$ relates to the diffusion coefficient via $\sigma=\sqrt{2D}$, $H$ is the Hurst parameter ($H=\alpha/2$) and $B_t^H$ - a continuous-time Gaussian process that starts at zero, has expectation zero  and has the following covariance function:
\begin{equation}
\mathrm{E}\left( B^H_t B^H_s\right) = \frac{1}{2}\left( |t|^{2H}+ |s|^{2H} - |t-s|^{2H} \right).
\end{equation}
For $H<\frac{1}{2}$ (i.e. $\alpha<1$), FBM produces subdiffusive trajectories. It corresponds to a scenario, in which a particle is hindered by mobile or immobile obstacles~\cite{JEO11}. It reduces to the free diffusion at $H=\frac{1}{2}$. And for $H>\frac{1}{2}$, FBM generates superdiffusive motion (Fig.~\ref{fig:example_traj}a).

The directed Brownian motion, also known as the diffusion with drift, is the solution to
\begin{equation}
dX_t^i = v_i dt+\sigma dB_t^{1/2,i},~~i=1,2,
\end{equation}  
where $v=(v_1,v_2)\in \mathbf{R}^2$  is the drift parameter. This process generates superdiffusion related to an active transport of particles driven by molecular motors. The velocity of the motors is modeled by the parameter $v$ (Fig.~\ref{fig:example_traj}b). For $v=0$, the process reduces to normal diffusion.

The Ornstein-Uhlenbeck process is known to model confined diffusion, which is a subclass of subdiffusion (Fig.~\ref{fig:example_traj}b). It corresponds to a particle inside a potential well and is a solution to the following stochastic differential equation:
\begin{equation}
dX_t^i = -\lambda_i(X_t^i-\theta_i)dt+\sigma dB_t^{1/2,i},~~i=1,2,~~\theta_i\in \mathbf{R}.
\end{equation}
Here, $\theta=(\theta_1,\theta_2)$ is the equilibrium position of the particle and $\lambda_i$ measures the strength of interaction. For $\lambda_i=0$, OU reduces to normal diffusion as well.

\section{Our dataset}
\label{sec:data}

\subsection{Real SPT data}
\label{sec:real}

The classifiers built in this study will be applied to the data from single-particle tracking experiment on G protein-coupled receptors and G proteins, already analyzed in Refs.~\cite{SUN17,WER19}. The receptors are of great interest, because they mediate the biological effects of many hormones and neourotransmitters and are also important as pharmacological targets~\cite{PIE02}. Their signals are transmitted to the cell interior via interactions with G proteins. The analysis of the dynamics of these two types of molecules  will shed more light on how the receptors and G proteins meet, interact and couple.

\begin{figure}
\includegraphics[scale=0.5]{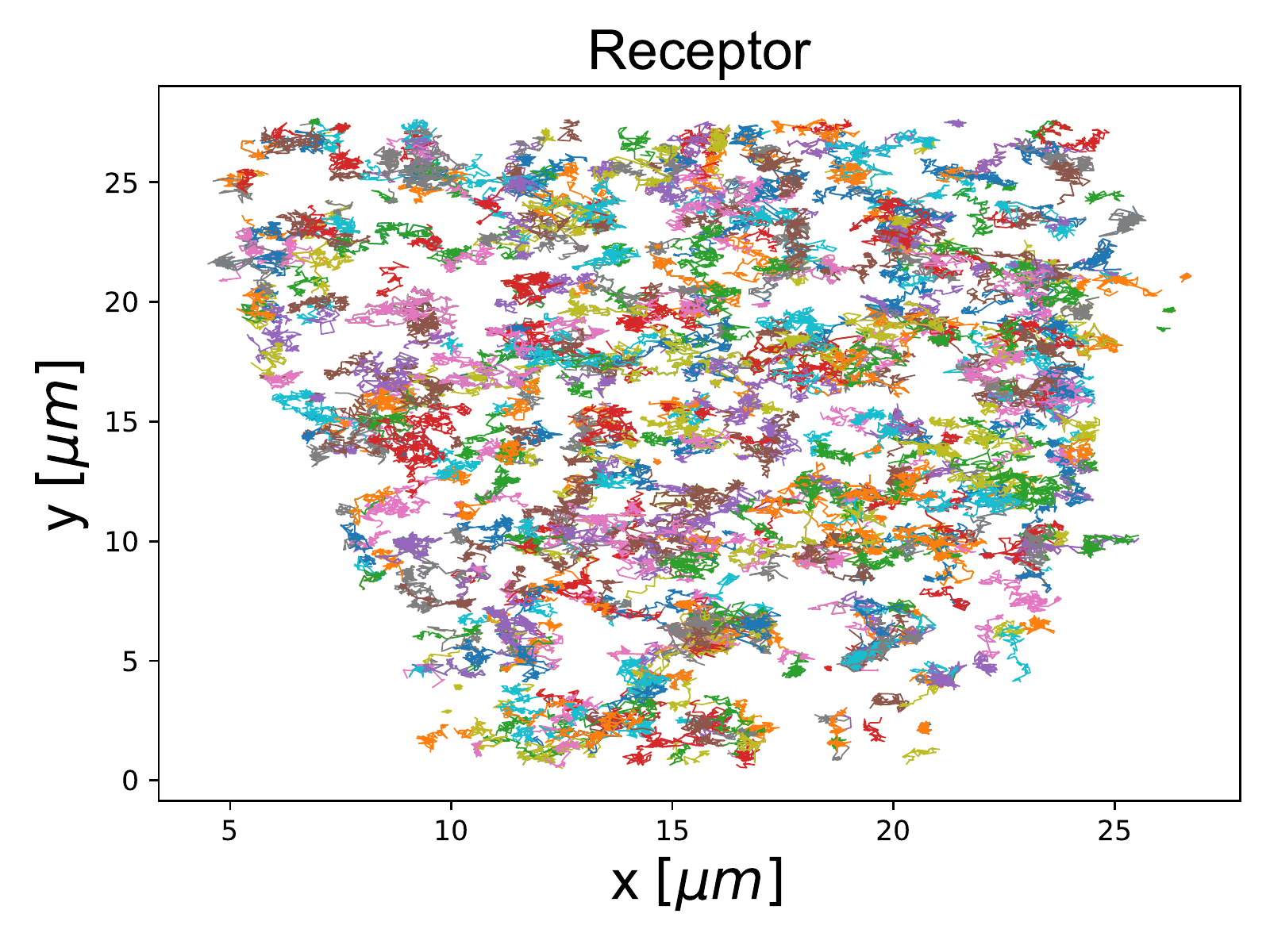}
\includegraphics[scale=0.5]{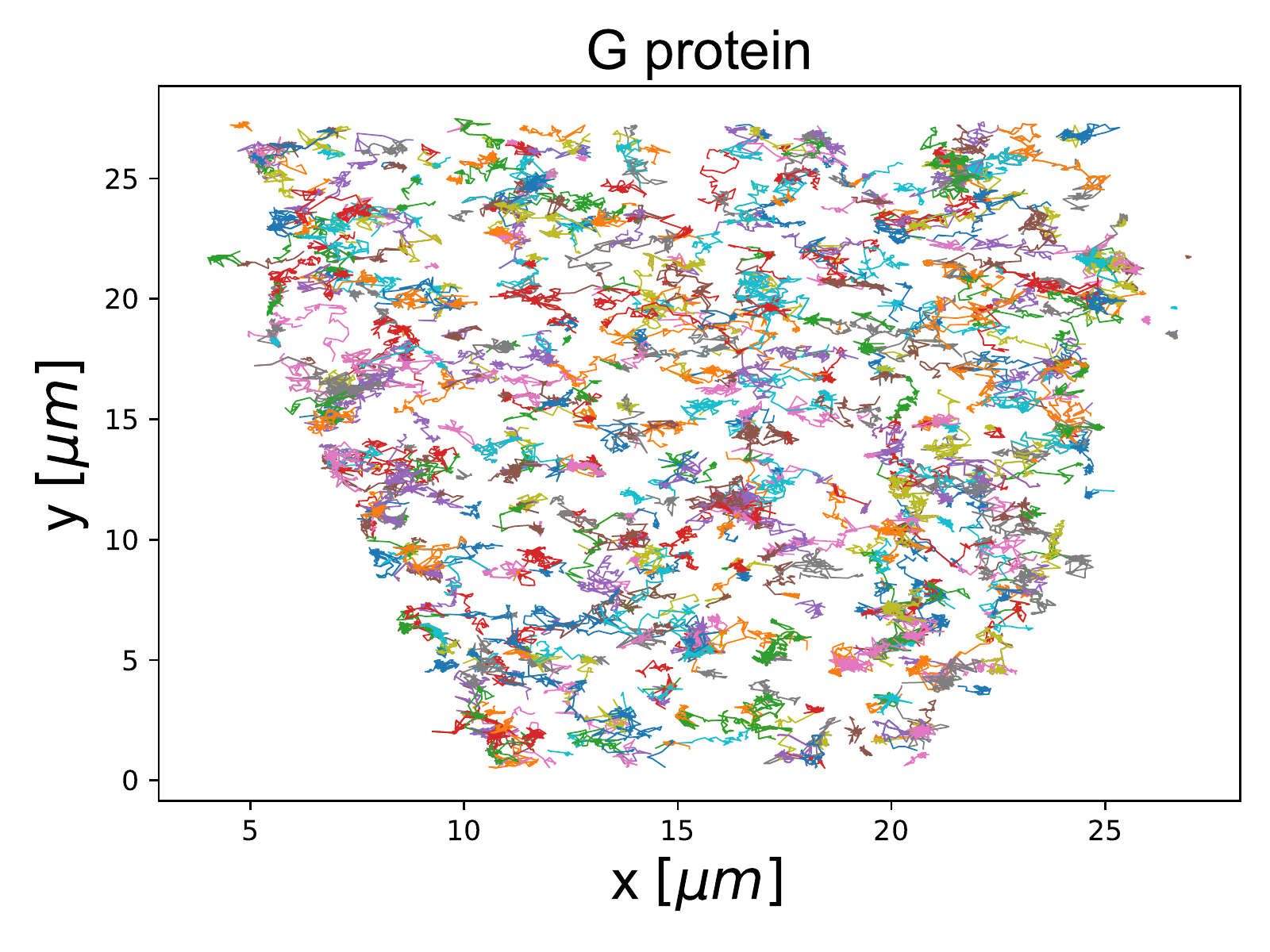}
\caption{(Color online) Trajectories of the receptors (left) and G proteins (right) used as input for the classifiers. Different colors are introduced to indicate different trajectories. The set of the receptors contains $1218$ trajectories and the one of G proteins -- $1037$ trajectories. The lengths of the trajectories are from range $[50, 401]$, the time step is equal to $28.4$ ms, and recorded positions are given in $\mu\text{m}$.\label{fig:real}}
\end{figure}

A subset of that data has been already studied by means of statistical methods in Ref.~\cite{WER19}. Since we are interested in using those results as a benchmark for our classifiers, we will focus on the very same subset of data in our analysis. Hence, only trajectories with at least 50 steps will be taken into account, resulting in 1037  G proteins and 1218 receptors. The trajectories under consideration for both types of molecules are visualized in Fig.~\ref{fig:real}.

\subsection{Synthetic data}
\label{sec:synthetic}

Building a classifier requires training data, which consists of a set of training examples~\cite{RAS15}. Each of these examples is a pair of an input (trajectory) and its output label (diffusion type). In an optimal scenario the training set would contain real trajectories with their true labels from e.g. previous experiments on the same type of cells. However, collecting a training set consisting of real trajectories is practically impossible. First of all, independently of the method used for analysis, the labels of such trajectories are affected by some uncertainties~\cite{WER19}. Moreover, typical machine learning algorithms require thousands of training examples to provide a reasonable function that maps an input to an output and can be used for classification of new input data. That is why one usually resorts to synthetic, computer generated trajectories to prepare the training set. In this case the true label of each trajectory is known in advance and it is rather cheap to generate many of them.

The stochastic processes described in Sec.~\ref{sec:stochastic} will be used to generate the training set. Just to recall, a discrete trajectory of a particle is given by 
\begin{equation}
X_n = \left( X_{t_0},X_{t_1},\dots,X_{t_N}\right),
\label{eq:traj}
\end{equation}
where $X_{t_i}=\left(X^1_{t_i},X^2_{t_i}\right)\in \mathbf{R}^2$ is the position of the particle at time $t_i = t_0+i\Delta t$, $i=0,1,\dots,N$. The lag $\Delta t$ between two consecutive observations is assumed to be constant. In tracking experiments, it is determined by the temporal resolution of the imaging method. However, we will assume the lag being equal to $1\,s$ in the simulations. Similarly, we will use $\sigma=1\,\mu m\, s^{-1/2}$ most of the time (see Sec.~\ref{sec: synthetic small d} for an exception to this choice). In total, $120 000$ trajectories have been generated for the main training set. Their length was randomly chosen from the range between 50 and 500 steps. No additional noise was added to the raw data in this set (see Sec.~\ref{sec: synthetic with noise} for a set with noise). 

A summary of the training set is presented in Table~\ref{tab:training}. The case of the free diffusion requires probably a short explanation. From the description in Sec.~\ref{sec:stochastic} we know that all of the models reduce to the normal Brownian motion for some specific values of the parametres ($H=0.5$ for FBM, $v=0$ for DBM and $\lambda=0$ for OU). However, it is very difficult to distinguish  anomalous diffusion processes from the normal one already if their parameters are in the vicinity of those values. That is why we extended the ranges of parameter values corresponding to the free diffusion. Although introduced here at a different level, this approach resembles the cutoff $c$ used in Ref.~\cite{WER19} to classify the results. As for the value, we take the smallest one considered therein.
\begin{table}
\begin{tabular}{c|c|c|c}
\hline \hline
Type of diffusion & Model & Parameter ranges & \# of trajectories  \\ 
\hline 
Normal diffusion & FBM & $H\in [0.5-c,0.5+c]$ & 20000  \\ 
                 & DBM & $v\in [0,c]$   & 10000  \\ 
                 & OU  & $\theta=0, \lambda=[0,c]$ & 10000  \\ 
\hline 
Subdiffusion     & FBM & $H\in [0.1,0.5-c)$ & 20000  \\ 
                 & OU  & $\theta=0, \lambda=(c,1]$ & 20000  \\ 
\hline 
Superdiffusion   & FBM & $H\in (0.5+c,0.9]$ & 20000  \\ 
                 & DBM & $v\in (c,1]$ & 20000  \\ 
\hline\hline 
\end{tabular} 
\caption{Summary of the synthetic trajectories used as the training set. The parameter $c$ was set to 0.1 in all simulations. If not specified otherwise, $\sigma=1\,\mu m\, s^{-1/2}$ and $\Delta t =1\,s$ were used.\label{tab:training}}
\end{table}

The Python package \texttt{fbm}~\cite{FLY17} was used to simulate the FBM trajectories as well as the Brownian motion part of the diffusion with drift. 
By default, the \texttt{fbm()} function from that package utilizes the Davies-Harte method~\cite{DAV87} for fastest performance. However, the method is known to fail for the Hurst parameter close to 1. If this occurs, the function fallbacks to the Hosking's algorithm~\cite{HOS84}. The OU process was generated with the \texttt{OrnsteinUhlenbeckProcess} object from the \texttt{stochastic} package~\cite{FLY18}.  This object uses the Euler-Maruyama method~\cite{KLO92} to produce realizations of the process.

\subsection{Adding noise}
\label{sec: synthetic with noise}

The synthetic dataset introduced in the previous section constitutes our main training set for building the classifiers. For the sake of comparison with statistical methods presented in Ref.~\cite{WER19}, it consists of pure trajectories, that do not suffer from any localization errors.

However, real data is usually affected by different kinds of noise. For instance, slow drift currents in the cytoplasm may induce low frequency noise. Typically, it may be reduced by various detrending methods~\cite{MIC10,MEL16}. In contrast, high frequency noise can be due to a variety of reasons: mechanical vibrations of the instrumental setup; particle displacement while the camera shutter is open; noisy estimation of true position from the pixelated microscopy image; error-prone tracking of particle positions when they are out of the camera focal plane~\cite{DES14,WEI13,BUR15,CAL16,WEI19}.

To account for different localization errors and to check their impact on the performance of the classifiers, we prepared a second training set by simply adding a normal Gaussian noise with zero mean and standard deviation $\sigma_{gn}$ to each simulated trajectory. We followed the procedure already used in Refs.~\cite{WAG17,KOW19}. That means, instead of setting  $\sigma_{gn}$ directly, we first introduced the signal to noise ratio,
\begin{equation}
Q=\left\{
\begin{array}{cc}
\frac{\sqrt{D\Delta t +v^2\Delta t^2}}{\sigma_{gn}} & {\rm for~DBM}, \\ 
\frac{\sqrt{D\Delta t }}{\sigma_{gn}} & {\rm otherwise},
\end{array} 
\right.
\label{eq:q}
\end{equation}
where $v=\sqrt{v_1^2 + v_2^2}$. For each trajectory, $Q$ was randomly set in the range from 1 (high noise) to 9 (low one). Then, Eq.~(\ref{eq:q}) was used to determine $\sigma_{gn}$ for given $D$ and $\Delta t$.

\subsection{Auxiliary training set}
\label{sec: synthetic small d}

During our first attempt to apply the classifiers to experimental data it turned out that the parameter $\sigma=1$, taken from Ref.~\cite{WER19}, may not be the best choice for real trajectories under investigation. Thus, we also prepared an auxiliary training set of synthetic trajectories, which were simulated with no noise and $\sigma=0.38$. This particular value of $\sigma$ corresponds to the diffusion coefficient $D=0.0715\,\mu m^2\,s^{-1}$ and will be explained in Sec.~\ref{sec: role of d}. All other parameters of the set are exactly the same as in the main training set introduced in Sec.~\ref{sec:synthetic}.

\section{Classification features}
\label{sec:feature}

The random forest and gradient boosting algorithms  require human-engineered features representing the trajectories for both the model training and the classification of new data. Choosing the right features constitutes a challenge and is crucial for the classification results. For instance, in Ref.~\cite{KOW19}  we considered a set of features proposed for the first time by Wagner et al~\cite{WAG17}. Although we did not apply them to real data, we showed that classifiers using those features do not generalize well to data generated with models different from the ones used for training.

A more detailed discussion on the role of the features will be addressed in a forthcoming paper. In this work we use a new set of features motivated by the statistical analysis carried out in Ref.~\cite{WER19}. The main conclusion of that paper was that, even though statistical methods going beyond the standard MSD classification may provide good results even for short trajectories, no method was found to be superior in all examples and one should actually combine different approaches to get reliable results.

Following this recommendation, we decided to extract features from all methods considered in Ref.~\cite{WER19} and to use them simultaneously as the input for our classifiers. Thus, our feature set will consist of:
\begin{itemize}
\item anomalous $\alpha$ exponent (fitted to TAMSD),
\item the diffusion coefficient $D$ (fitted to TAMSD),
\item the standarized value
\begin{equation}
T_N = \frac{D_N}{\sqrt{\hat{\sigma}_N^2 (t_N -t_0)}}
\end{equation}
of the maximum distance $D_N$ traveled by a particle,
\begin{equation}
D_N = \max_{i=1,2,\dots,N}\Vert X_{t_i}-X_{t_0}\Vert,
\end{equation}
where $\hat{\sigma}_N$ is a consistent estimator of the standard deviation of $D_N$,
\begin{equation}
    \hat{\sigma}_{N}^2 = \frac{1}{2N\Delta t}\sum^N_{j=1} \Vert X_{t_{j}}-X_{t_{j-1}} \Vert^2_2,
\end{equation}
\item the power $\gamma^p$ (in the function $kn^{\gamma^p}$) fitted to $p$-variation~\cite{MAG09,BUR10}
\begin{equation}
\hat{V}_n^{(p)}=\sum_{k=0}^{N/n - 1}\Vert X_{(k+1)n}-X_{kn}\Vert^p
\end{equation}
for values of $p$ from 1 to 5.
\end{itemize}
Note that the first two of the above features were included in the feature set used in Refs.~\cite{KOW19}. To determine their values, the maximum lag equal to 10\% of each trajectory's length was used to calculate the corresponding TAMSD curve.

\section{Results}
\label{sec:results}

We used the \texttt{scikit-learn}~\cite{PED11} implementations of the random forest and gradient boosting algorithms. As already stated in Ref.~\cite{KOW19}, a cluster of 24 CPUs with 25 GB total memory was used to perform the computation. The processing time (feature extraction, hyperparameter tuning, training and validation of a model) was of the order of two hours in each case. If not stated otherwise, the dataset without noise (see Sec.~\ref{sec:synthetic}) was used to train the classifiers.

\subsection{Details of the classifiers}
\label{sec:details}

In order to find optimal models, we used the \texttt{RandomisedSearchCV} method from \texttt{scikit-learn} library. It allows to perform a search over a grid of hyperparameter ranges. Here, a hyperparameter of the model is understood as a parameter, whose value is set before the learning process begins (it cannot be derived simply by training of the model).

In Table.~\ref{tab:hyperparameters}, the optimal values of the hyperparameters for our training set are listed.  The ``with $D$'' column in the table refers to the full set of features defined in Sec.~\ref{sec:feature}. The ``no $D$'' columns corresponds to a reduced feature set with the diffusion coefficient $D$ removed from consideration. The reason for introducing the latter set will be explained in Sec.~\ref{sec:results_real}.
\begin{table}
\begin{tabular}{l||c|c|c|c}
\hline\hline
 & \multicolumn{2}{c|}{Random forest} & \multicolumn{2}{c}{Gradient boosting} \\ 
\cline{2-5}
 & with $D$ & no $D$ & with $D$ & no $D$ \\ 
\hline 
\texttt{bootstrap} & $True$ & $True$ & NA & NA \\ 
\hline 
\texttt{criterion} & $entropy$ & $entropy$ & NA & NA \\ 
\hline 
\texttt{max\_depth} & 60 & 10 & 10 & 10 \\ 
\hline 
\texttt{max\_features} & $log_2$ & $sqrt$ & $log_2$ & $log_2$ \\ 
\hline 
\texttt{min\_samples\_leaf } & 4 & 2 & 2 & 2 \\ 
\hline 
\texttt{min\_samples\_split} & 2 & 10 & 2 & 2 \\ 
\hline 
\texttt{n\_estimators} & 900 & 600 & 100 & 100 \\ 
\hline\hline
\end{tabular} 
\caption{Hyperparameters of the optimal classifiers found with both methods. Their meaning is explained in Sec.~\ref{sec:details}. The ``with $D$'' column refers to the full feature set, ``no $D$'' one - to the feature set after removal of the diffusion coefficient $D$. $NA$ stands for ``Not Applicable'' (the first two parameters are random forest specific).  \label{tab:hyperparameters}}
\end{table}
The \texttt{bootstrap} hyperparameter is a boolean value. It decides whether bootstrap samples ($True$) or the whole data set ($False$) are used to build each single tree. \texttt{Criterion} specifies, which function should be used to measure the quality of a split of data into subsamples at a new node of the tree. Gini impurity and information entropy are available for that purpose~\cite{RAS15}. The \texttt{max\_depth} is the maximum depth (the number of levels) of each decision tree.  The number of features to consider when looking for the best split is given by \texttt{max\_features}. If equal to \texttt{log2} (\texttt{sqrt}), then the number is calculated as the logarithm (square root) of the number of features.  \texttt{Min\_samples\_split} specifies the minimum number of samples required to split a subset of data at an internal node of the tree. \texttt{Min\_samples\_leaf} is the minimum number of samples required to be at a leaf node (a node representing a class label). Finally, \texttt{n\_estimators} gives the number of trees in the ensemble.

As it follows from Table~\ref{tab:hyperparameters}, the ensemble found with gradient boosting is significantly smaller than the one generated with the random forest method.   

\subsection{Performance of the classifiers}

Since our synthetic data set is perfectly balanced (same number of trajectories in each class), we may start the analysis of the classifiers simply by looking at their accuracy. It is one of the basic measures to assess the classification performance, defined as the number of correct predictions divided by the total number of preditions.

From the results listed in Table~\ref{tab:accuracy} it follows that in the case of the training set, the gradient boosting method is the more accurate one, even though the differences are small. Moreover, its decline in accuracy after the removal of $D$ from the feature set is smaller than for random forest. However, the latter one performs a little bit better on the test set, indicating a small tendency of GB to overfit. Despite these differences, both classifiers perform very well.
\begin{table}
\begin{tabular}{c||c|c||c|c}
\hline\hline 
  & \multicolumn{2}{c||}{Random forest} & \multicolumn{2}{c}{Gradient boosting}  \\ 
\cline{2-5} 
Data set & with $D$ & no $D$ & with $D$ & no $D$ \\ 
\hline\hline 
Training & 0.977 & 0.953 & 0.992 & 0.989 \\ 
\hline 
Test & 0.948 & 0.946 & 0.947 & 0.944 \\ 
\hline \hline
\end{tabular} 
\caption{Accuracies of the optimal classifiers for both the training and the test data.\label{tab:accuracy}}
\end{table}

\begin{figure}
	\includegraphics[scale=0.5]{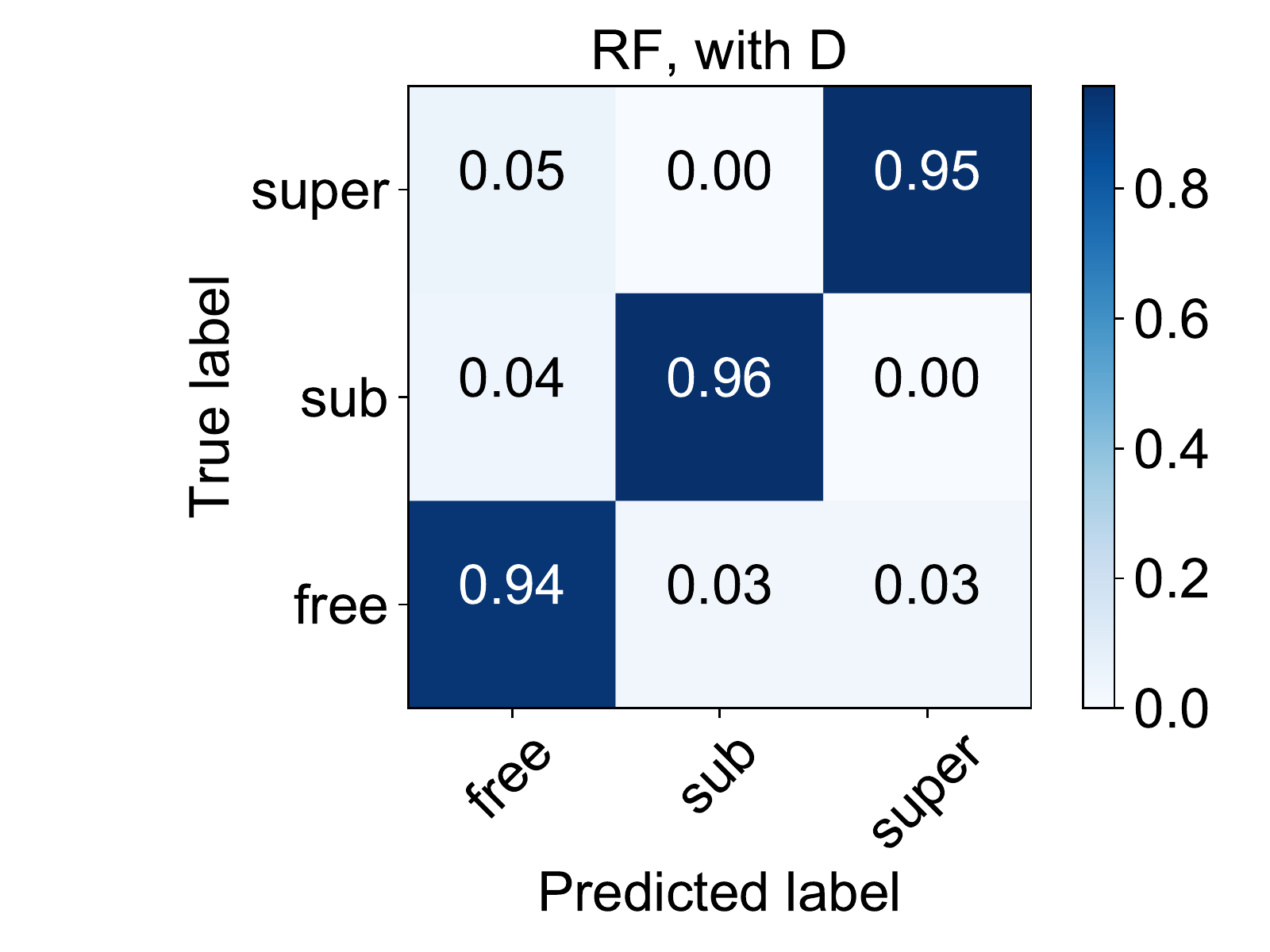}
	\includegraphics[scale=0.5]{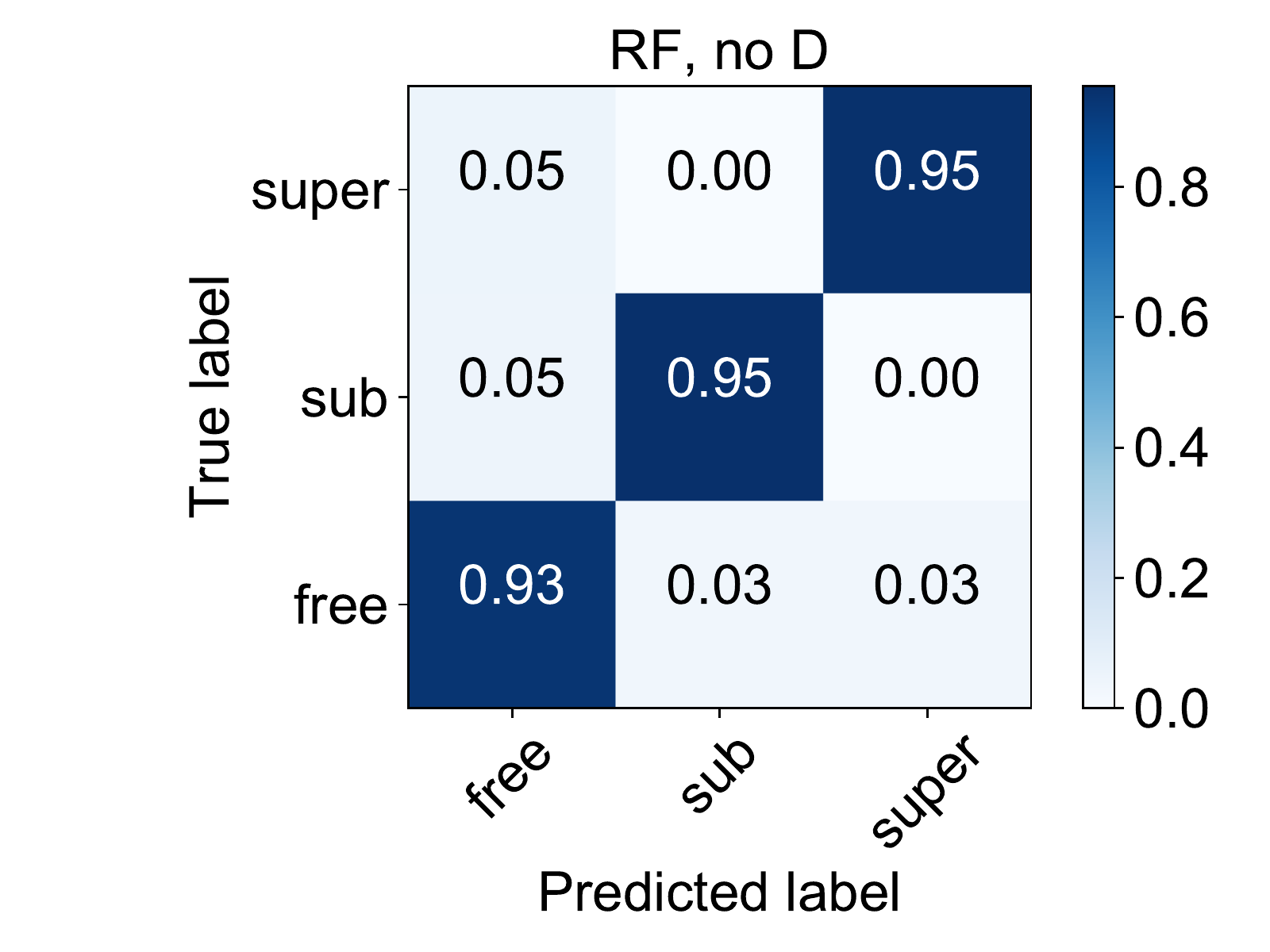}\\
	\includegraphics[scale=0.5]{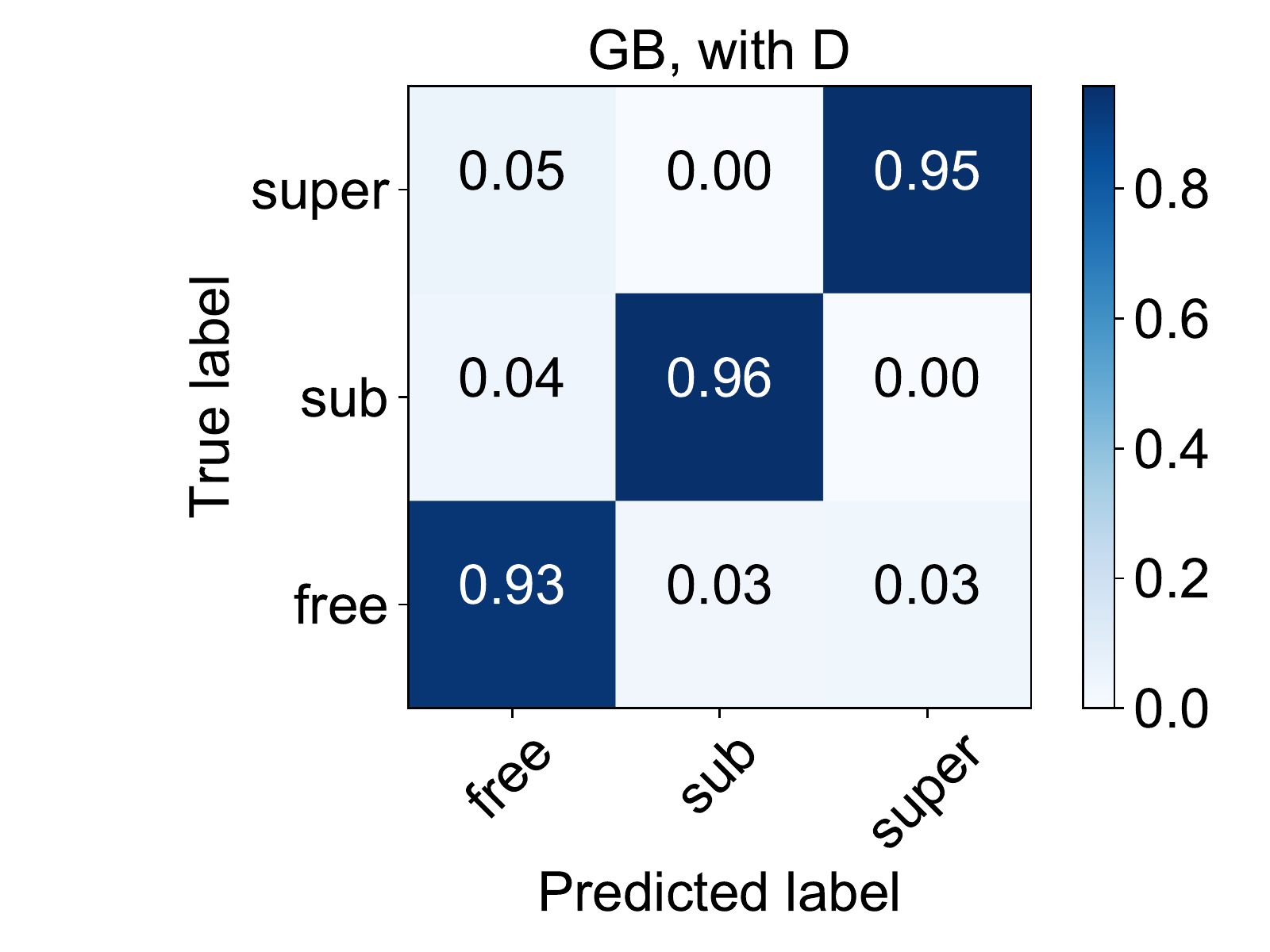}
	\includegraphics[scale=0.5]{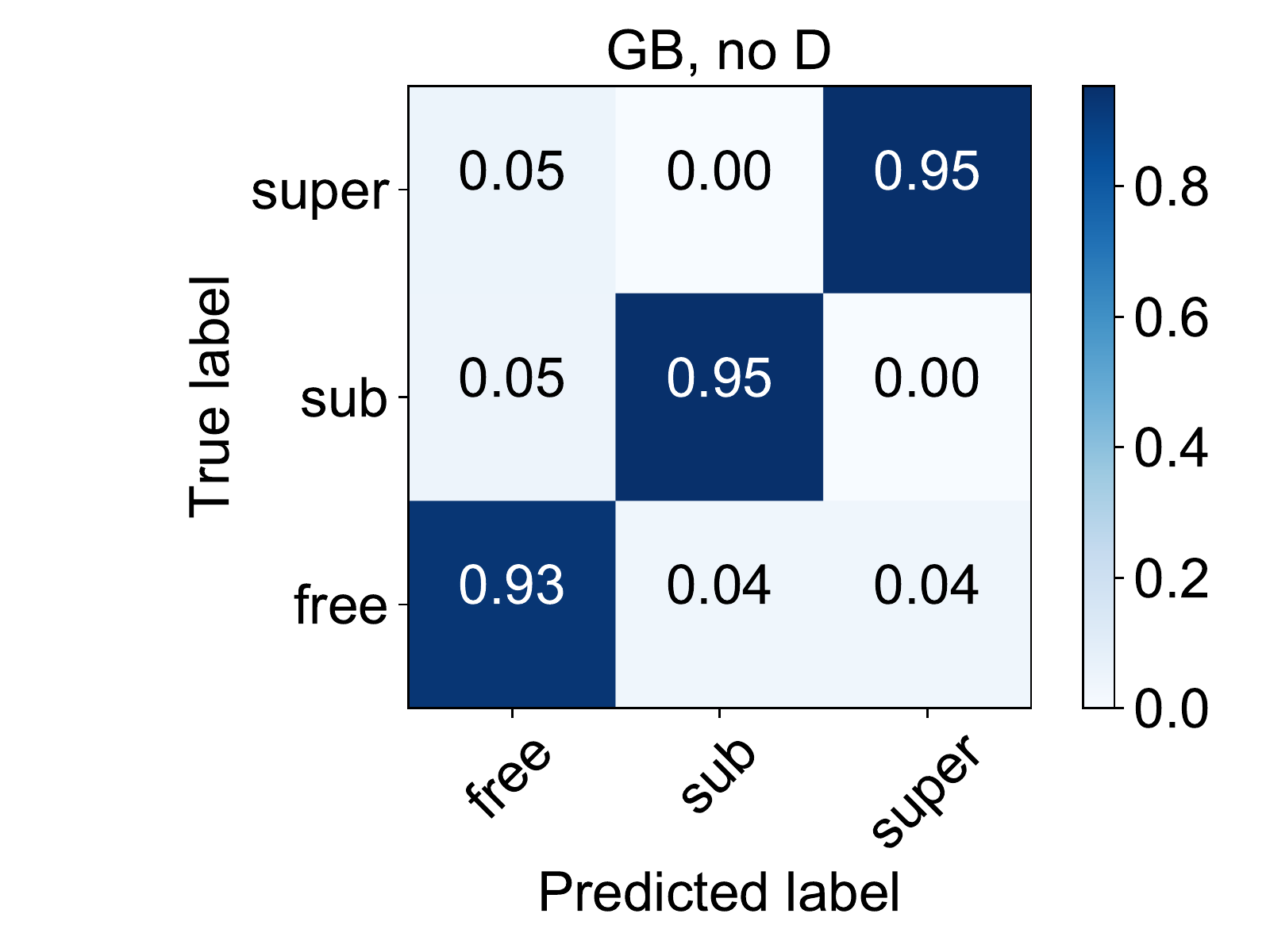}
\caption{Normalized confusion matrices for random forest and gradient boosting classifiers. The ``with $D$'' label refers to the full feature set, ``no $D$'' one - to the feature set after removal of the diffusion coefficient $D$. All results are rounded to two decimal digits.\label{fig:confusion}}	
\end{figure}

The normalized confusion matrices of the classifiers are presented in Fig.~\ref{fig:confusion}. By definition, an element $C_{ij}$ of the confusion matrix is equal to the number of observations known to be in class $i$ (true labels) and predicted to be in class $j$ (predicted labels)~\cite{RAS15}. In all cases, the worst performance (93\% of correctly predicted labels) is observed for the normal diffusion. This relates probably to the fact that in our synthetic training set we also tagged  anomalous trajectories with parameters slightly deviating from the normal ones as free diffusion.  

The data collected in Fig.~\ref{fig:confusion} may be used to calculate some other popular measures giving more insight into the performance of the classifiers: precision, recall and F1 score~\cite{PER55}. Precision is the fraction of correct predictions of a class among all predictions of that class. It tells us how often a classifier is correct if it predicts a given class. Recall is the fraction of correct predictions of a given class over the total number of members of that class. It measures the number of relevant results within a predicted class. A harmonic mean of precision and recall gives the F1 score - another measure of classifier's accuracy.

As we see from Table~\ref{tab:measures}, both models return much more relevant results than the irrelevant ones (high precision). Moreover, they yield most of the relevant results (high recall). The F1 scores resemble the accuracies given in Table~\ref{tab:accuracy}.

\begin{table}
\begin{tabular}{c|c|c||c|c|c||c}
\hline \hline
Method & Features & Measure & Normal diffusion & Subdiffusion & Superdiffusion & Total/Average \\ 
\hline \hline
 &  & Support & 12000 & 12000 & 12000 & 36000 \\ 
\cline{3-7}
 &  & Precision & 0.912 & 0.969 & 0.966 & 0.949 \\ 
\cline{3-7} 
 & with $D$ & Recall & 0.935 & 0.958 & 0.951 & 0.948 \\ 
\cline{3-7}  
RF &  & F1 & 0.923 & 0.963 & 0.959 & 0.948 \\ 
\cline{2-7} 
 &  & Support & 12000 & 12000 & 12000 & 36000 \\ 
\cline{3-7}
 &  no $D$ & Precision & 0.908 & 0.967 & 0.967 & 0.947 \\ 
\cline{3-7} 
 &  & Recall & 0.935 & 0.955 & 0.950 & 0.947 \\ 
\cline{3-7}
 &  & F1 & 0.921 & 0.961 & 0.958 & 0.947 \\ 
\hline \hline
 &  & Support & 12000 & 12000 & 12000 & 36000 \\ 
\cline{3-7}
 &  & Precision & 0.911 & 0.967 & 0.965 & 0.948 \\ 
\cline{3-7} 
 & with $D$ & Recall & 0.933 & 0.958 & 0.951 & 0.947 \\ 
\cline{3-7}  
GB &  & F1 & 0.922 & 0.962 & 0.958 & 0.947 \\ 
\cline{2-7} 
 &  & Support & 12000 & 12000 & 12000 & 36000 \\ 
\cline{3-7}
 &  no $D$ & Precision & 0.907 & 0.963 & 0.964 & 0.945 \\ 
\cline{3-7} 
 &  & Recall & 0.928 & 0.954 & 0.951 & 0.944 \\ 
\cline{3-7}
 &  & F1 & 0.917 & 0.958 & 0.957 & 0.944 \\ 
\hline \hline
\end{tabular} 
\caption{Detailed performance analysis of both classification methods on the test data. Support is  the number of trajectories known to belong to a given class. All results are rounded to two decimal digits.\label{tab:measures}}
\end{table}

\subsection{Feature importances}

While working with the human-engineered features, it may happen that some of them are more informative than the others. Therefore, knowing the relative importances of the features is useful, because it can provide further insight into the data and the classification model. The features with high importances are the drivers of the outcome. The least important ones might often be omitted from the model, making it faster to fit and predict. The latter is of particular significance in case of models with very large feature sets, as it may additionally help to reduce the dimensionality of the problem.

There are several ways to determine the feature importances. The one implemented in the \texttt{scikit-learn} library is defined as the total decrease in node impurity caused by a given feature, averaged over all trees in the ensemble~\cite{BRE84}. In other words, the Gini impurities~(\ref{eq:gini}) are calculated before and after each split on a given feature to determine the total decrease in the impurity related to that feature. The outcome is then averaged over all trees in the ensemble.

Relative feature importances for both classifiers are shown in Table~\ref{tab:importances}. The features are ordered according to the descending scores in case of the random forest with $D$. 
\begin{table}
\begin{tabular}{c||c|c||c|c}
\hline \hline
 & \multicolumn{2}{c||}{Random forest} & \multicolumn{2}{c}{Gradient boosting}  \\ 
\cline{2-5} 
Feature & with $D$ & no $D$ & with $D$ & no $D$ \\ 
\hline \hline
$2$-var  & \textbf{0.296} & \textbf{0.239} & 0.238 & 0.160 \\ 
\hline 
$\alpha$  & 0.201 & 0.197 & \textbf{0.274} & 0.125 \\ 
\hline 
$3$-var & 0.178 & 0.183 & 0.108 & \textbf{0.245} \\ 
\hline 
$1$-var & 0.171 & 0.200 & 0.204 & 0.210 \\ 
\hline 
$4$-var & 0.078 & 0.110 & 0.095 & 0.145 \\ 
\hline 
$T_N$ & 0.038 & \underline{0.032} & 0.030 & \underline{0.037} \\ 
\hline 
$5$-var & 0.022 & 0.038 & 0.034 & 0.077 \\ 
\hline 
$D$ & \underline{0.017} & -- & \underline{0.016} & -- \\ 
\hline \hline
\end{tabular} 
\caption{Feature importances for both methods, sorted in the descending order of the scores in case of random forest with $D$. The bold face indicates the most important features in each case. The least important ones are underlined. The ``with $D$'' label refers to the full feature set, ``no $D$'' one - to the feature set after removal of the diffusion coefficient $D$.  \label{tab:importances}}
\end{table}
We see that the $p$-variation for $p=2$ ($2$-var) is the most informative feature, followed by the anomalous exponent $\alpha$. The diffusion coefficient is the least important feature. After its removal the relative importances of the remaining features changed. The differences between them are smaller now.  Moreover, the $1$-var became the second most important attribute and $T_N$ - the least important one.  

Gradient boosting differs slightly from the random forest. In case with $D$, the order of the top two features is reversed. After removal of $D$, $3$-var and $1$-var  became the most informative ones. The exponent $\alpha$  lost much of its importance. And again, $T_N$ is the least informative attribute.

\subsection{Note on auxiliary classifiers}

Apart from  the main collection of synthetic trajectories described in Sec.~\ref{sec:synthetic}, we generated two additional training sets. The first one was built from the main set by simply adding noise (Sec.~\ref{sec: synthetic with noise}) and the second one --  with a smaller value (0.38 vs 1) of the parameter $\sigma$ (Sec.~\ref{sec: synthetic small d}).

Those sets were then used to train new classifiers. Their accuracies are listed in Table~\ref{tab:accuracy new}.
\begin{table}
\begin{tabular}{c||c|c||c|c}
\hline\hline 
  & \multicolumn{2}{c||}{Random forest} & \multicolumn{2}{c}{Gradient boosting}  \\ 
\cline{2-5} 
Data set & with $D$ & no $D$ & with $D$ & no $D$ \\ 
\hline\hline 
with noise & 0.946 & 0.932 & 0.946 & 0.930 \\ 
\hline
with $\sigma=0.38$ & 0.953 & 0.950 & 0.952 & 0.950 \\ 
\hline \hline
\end{tabular} 
\caption{Accuracies of the classifiers trained on auxiliary data sets: the first with noise (see Sec.~\ref{sec: synthetic with noise}) and the second with  $\sigma=0.38$ (Sec.~\ref{sec: synthetic small d}) ).\label{tab:accuracy new}}
\end{table}
Note that those values are very similar to the ones presented in Table~\ref{tab:accuracy}. Thus, all classifiers perform very well on their corresponding synthetic test sets. Interestingly, the machine learning algorithms seem to deal excellent with noisy data, as there is no significant drop  in the accuracy of the classifiers trained on that data.

The basic characteristics of the additional classifiers turned out to be practically indistinguishable from the ones presented in the previous sections. Thus, we will skip their detailed description for the sake of readability.

\subsection{Application to real data}
\label{sec:results_real}

\subsubsection{Summary of statistical methods}

In Table~\ref{tab:statres}, classification results from Ref.~\cite{WER19} for the G protein-coupled receptors and G proteins  (see Sec.~\ref{sec:real} for details)  are summarized.
\begin{table}
\begin{tabular}{c||c|c||c|c||c|c||c|c||c|c}
\hline\hline 
 & \multicolumn{2}{c||}{MSD}  & \multicolumn{2}{c||}{MSD test} & \multicolumn{2}{c||}{MAX} & \multicolumn{2}{c||}{1-var} & \multicolumn{2}{c}{2-var} \\ 
\cline{2-11}
 & R & G & R & G & R & G  & R & G & R & G \\ 
\hline \hline
Normal diffusion & 19\% & 22\% & 79\% & 76\% & 79\% & 76\% & 53\% & 52\% & 47\% & 51\% \\ 
\hline 
Subdiffusion & 80\% & 72\% & 21\% & 24\% & 21\% & 24\% & 47\% & 46\% & 53\% & 48\% \\ 
\hline 
Superdiffusion & 1\% & 6\% & 0 \% & 1\% & 0\% & 1\% & 0\% & 1\% & 0\% & 2\% \\ 
\hline \hline
\end{tabular} 
\caption{Summary of the classification results from Ref.~\cite{WER19}. Columns labeled with R and G correspond to the G protein-coupled receptors and G proteins, respectively. The MSD data was calculated for $c=0.1$ (see Sec.~\ref{sec:synthetic} for explanation) and the maximum lag equal to 10\% of the trajectories' lengths. Due to rounding, the numbers may not add up precisely to 100\%. \label{tab:statres}}
\end{table}
Except the standard MSD method, the authors used statistical testing procedures based on: (a) MSD (referred as ``MSD test'' in Table~\ref{tab:statres}), (b) maximum distance traveled by a particle (``MAX'') and (c) $p$-variations at different values of $p$ (``1-var'' and ``2-var''). As wee see, the methods do not yield coherent results. MSD classifies most of the trajectories as subdiffusion. The MAX and MSD test procedures indicate a prevalence of freely diffusing particles in the same data set. The $p$-var tests give similar proportions of normal and subdiffusive trajectories. Moreover, only the standard MSD method is able to recognize a noticeable subset of trajectories as superdiffusion. Further analysis with synthetic data revealed that the $p$-var method is the most accurate one for FBM, while the MSD/MAX tests are the best choice (in terms of errors) for   OU and DBM processes.

\subsubsection{Classification with full feature set}

Our first attempt to classify the data with the whole feature set defined in Sec.~\ref{sec:feature} is presented in Table~\ref{tab:realclasd}.
\begin{table}
	\begin{tabular}{ c|| c| c || c| c } 
	    \hline\hline
		& \multicolumn{2}{c||}{Random forest}  & \multicolumn{2}{c}{Gradient boosting} \\
		\cline{2-5}
		      & Receptor & G protein & Receptor & G protein \\ \hline
		Normal diffusion 	  & 38\%     & 44\%      & 38\%     & 38\% \\ 
		Subdiffusion      & 61\%     & 54\%      & 60\%     & 55\% \\  
		Superdiffusion    & 0\%     & 1\%      & 0\%     & 5\%  \\  
		\hline\hline
	\end{tabular}
	\caption{Diffusion modes of real trajectories found with classifiers trained on the main synthetic dataset ($\sigma=1$, no noise; see Sec.~\ref{sec:synthetic}) with the full set of features (referred to as ``with D'' in the previous sections).  Due to rounding, the numbers may not add up precisely to 100\%.\label{tab:realclasd}}
\end{table}
As we see, both methods work similarly, with gradient boosting recognizing more G protein trajectories as a superdiffusive motion. Most of the trajectories are classified as normal or subdiffusion, with the prevalence of the latter. 
Note that the numbers given in Table~\ref{tab:realclasd} do not match any of the results generated with the statistical methods. Thus, it is really hard to judge which method should be chosen to work with real data.  

\subsubsection{Role of $D$ and classification with reduced feature set}
\label{sec: role of d}

\begin{figure}
\includegraphics[scale=1]{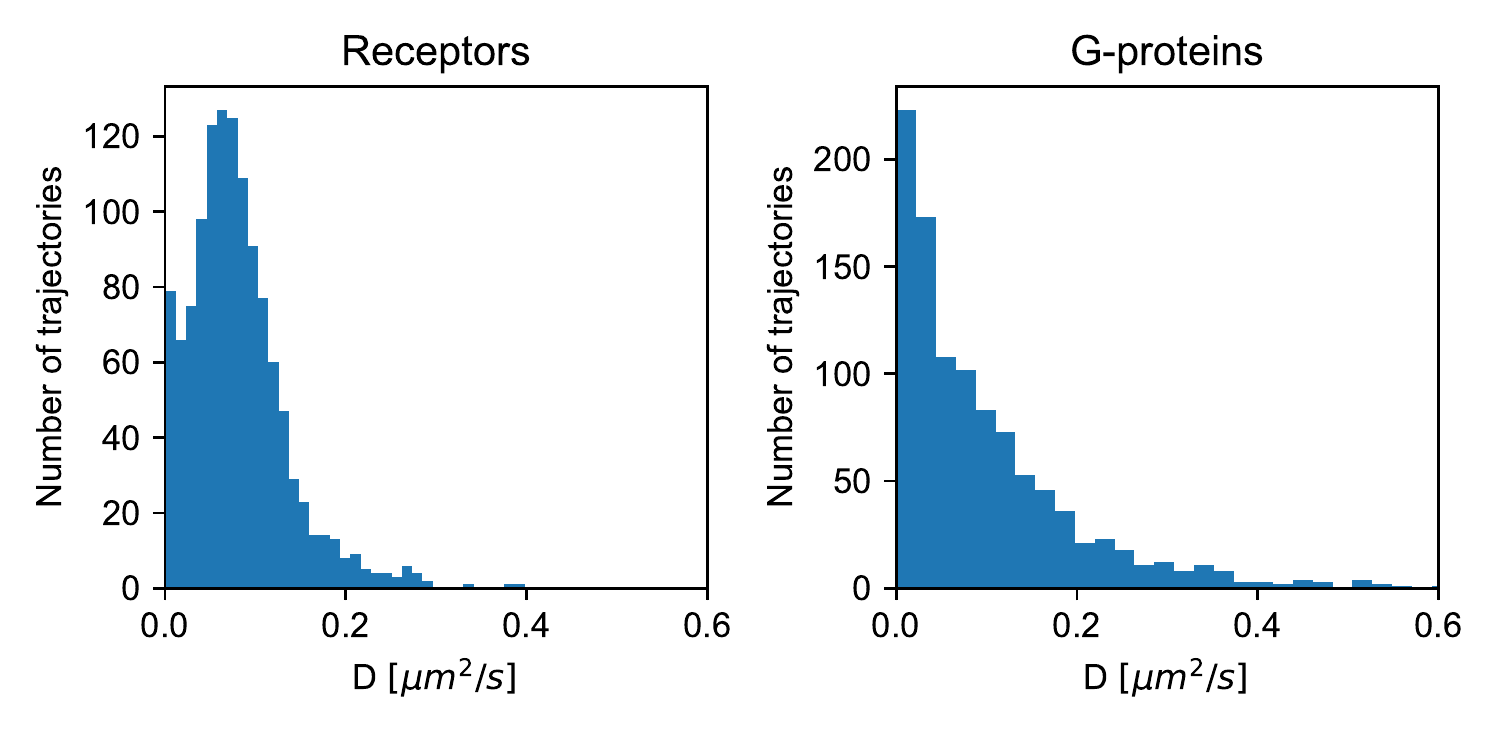}
\caption{Distribution of $D$ among trajectories in the real data set.\label{fig:hist}}
\end{figure}

To pin down a possible cause for the deviation from the statistical methods, let us have a look at the distribution of values of the generalized diffusion coefficient $D$ among the trajectories in the real data set. The corresponding histograms for  the G protein-coupled receptors and G proteins  are shown in Fig.~\ref{fig:hist}.
To calculate the histograms, $D$ was simply extracted from the MSD curves under the assumption of the anomalous diffusion model~\cite{WER19}. Its values in the data set turned out to be  much smaller than in the synthetic data set, with $D= 0.0715$ being the most frequent one. However, the synthetic training set was generated with  $\sigma=1$ (i.e. $D=0.5$) for all types of diffusion. Thus, the discrepancy in the classification results  from any of the methods presented in~\cite{WER19} may simple be caused by the fact, that the classifiers were trained for a different regime of diffusion. 

To check this hypothesis let us classify the trajectories with the models trained on the data set generated with $\sigma=0.38$, which corresponds to $D=0.0715$. Classification results with the full set of features are presented in Table~\ref{tab:realclasssmalld}.
\begin{table}
	\begin{tabular}{ c|| c| c || c| c } 
	    \hline\hline
		& \multicolumn{2}{c||}{Random forest}  & \multicolumn{2}{c}{Gradient boosting} \\
		\cline{2-5}
		      & Receptor & G protein & Receptor & G protein \\ \hline
		Normal diffusion 	  & 54\%     & 51\%      & 54\%     & 51\% \\ 
		Subdiffusion      & 45\%     & 47\%      & 45\%     & 46\% \\  
		Superdiffusion    & 0\%     & 1\%      & 0\%     & 1\%  \\  
		\hline\hline
	\end{tabular}
	\caption{Diffusion types found in real data with the classifiers trained on the auxiliary dataset ($\sigma=0.38$, no noise; see Sec.~\ref{sec: synthetic small d}) with the full set of features (referred to as ``with D'' in the previous sections). Due to rounding, the numbers may not add up precisely to 100\%.\label{tab:realclasssmalld}}
\end{table}
The results differ from the previous classification. Now, the shares of the trajectories are very similar to the ones obtained with $1$-var and $2$-var methods from Ref.~\cite{WER19}.  Most of the trajectories belong either to the normal diffusion or the subdiffusion class, with the first having a slightly larger count. There is only a bunch of data samples recognized as superdiffusion in case of G~proteins.
 
Treating the statistical results as a reference point, we can indeed say that adjusting $\sigma$ to the real data set improved the results. However, this procedure is not really convenient, as it requires generation of a new synthetic data set, extracting of features and training of a new classifier practically every time new experimental samples are arriving.

In search of a more universal procedure we decided to train new classifiers on the reduced set of features not containing the diffusion coefficient $D$, but we used the main synthetic set with $\sigma=1$ as in Ref.~\cite{WER19}. Since $D$ turned out to be the least informative feature (Table~\ref{tab:importances}), it was the natural candidate for the removal anyway.
\color{black}
 We know already, that the accuracy of the classifiers without $D$ is a little bit smaller (Table~\ref{tab:accuracy}). Nevertheless, we expect them to work better on unseen data.
Indeed, even though the choice of $\sigma$ was not optimal, the classification results shown in Table~\ref{tab:realclasnod} resemble the ones obtained above (Table~\ref{tab:realclasssmalld}) and the $1$-var and $2$-var methods from Ref.~\cite{WER19}.
\begin{table}
	\begin{tabular}{ c|| c| c || c| c } 
	    \hline\hline
		& \multicolumn{2}{c||}{Random forest}  & \multicolumn{2}{c}{Gradient boosting} \\
		\cline{2-5}
		      & Receptor & G protein & Receptor & G protein \\ \hline
		Normal diffusion 	  & 52\%     & 53\%      & 56\%     & 54\% \\ 
		Subdiffusion      & 46\%     & 45\%      & 43\%     & 43\% \\  
		Superdiffusion    & 0\%     & 1\%      & 0\%     & 1\%  \\  
		\hline\hline
	\end{tabular}
	\caption{Results of classifiers trained on the main synthetic dataset ($\sigma=1$, no noise) with the reduced set of features, after removal of $D$ (referred to as ``no D'' in the previous sections).  Due to rounding, the numbers may not add up precisely to 100\%.\label{tab:realclasnod}}
\end{table}
Again, most of the trajectories belong either to the normal diffusion or the subdiffusion class, with the first having a larger count. Just few data samples are recognized as superdiffusion in case of G proteins.

The advantage of the classification with the reduced data set over the one with the adjustment of $\sigma$ lies in that the classifier is trained only once and then may be simply applied to any unseen samples. It does not require a recurring and time consuming procedure of adjustment of $\sigma$, generation of tailor-made training data, extraction of features and training of the classifier every time a new set of experimental trajectories is arriving for analysis.

The agreement with the $p$-variation procedure for small values of $p$ makes perfect sense, if we recall that $2$-var and $1$-var belong to be the most informative among the features used by the classifiers to distinguish between the data samples.

\subsubsection{Impact of noise}

Although it goes beyond a comparison of machine learning algorithms with the statistical methods from Ref.~\cite{WER19}, as its authors used only pure trajectories, we would like to conclude this section by applying the classifier trained on noisy data (Sec.~\ref{sec: synthetic with noise}) to the real trajectories. Results of classification with the reduced feature set are shown in Table~\ref{tab:realclasnoise}. 

\begin{table}
	\begin{tabular}{ c|| c| c || c| c } 
	    \hline\hline
		& \multicolumn{2}{c||}{Random forest}  & \multicolumn{2}{c}{Gradient boosting} \\
		\cline{2-5}
		      & Receptor & G protein & Receptor & G protein \\ \hline
		Normal diffusion 	  & 62\%     & 58\%      & 63\%     & 58\% \\ 
		Subdiffusion      & 38\%     & 40\%      & 36\%     & 40\% \\  
		Superdiffusion    & 0\%     & 2\%      & 0\%     & 3\%  \\  
		\hline\hline
	\end{tabular}
	\caption{Results of classifiers trained on noisy data ($\sigma=1$) with the reduced set of features (referred to as ``no D'' in the previous sections).  Due to rounding, the numbers may not add up precisely to 100\%.\label{tab:realclasnoise}}
\end{table}

The introduction of noise has changed the results. Although still most of the trajectories belong either to the normal diffusion or to the subdiffusion class, the first has now larger count compared with the case without noise in the synthetic data. As we already pointed out in Sec.~\ref{sec:synthetic}, the boundary between the normal and anomalous modes in the vicinity of $\alpha=1$ is not particularly well defined even in the absence of noise. This boundary is further blurred in the presence of noise, resulting in the observed rearrangement of class memberships. However, if we still  would like to relate the ``noisy'' results with the ones from Ref.~\cite{WER19}, they are close to an average of $1$-var and $MAX$ methods.

\section{Discussion}

Machine learning methods used for classification of SPT data are known to sometimes fail to generalize to unseen data~\cite{KOW19}. In this paper, we revisited our ML approach to trajectory classification and presented a new set of features, which are required by the classifiers to process the input data. This new set allows the random forest and gradient boosting classifiers to transfer the knowledge from a synthetic training set to real data. The classifiers were tested on a subset of experimental data describing G proteins and G protein-coupled receptors~\cite{SUN17}. The results were then compared to four statistical testing procedures introduced in Ref.~\cite{WER19}.

We have shown that the choice of the feature set is crucial, as even a small change in its content may significantly impact the behavior of the classifiers. We decided to use a set consisting of  the anomalous $\alpha$ exponent, the diffusion coefficient $D$, the maximum distance traveled by a particle, the power $\gamma^p$ fitted to $p$-variation for values of $p$ from 1 to 5. These features were extracted from several statistical methods presented in Ref.~\cite{WER19}. Since none of the methods turned out to be superior to the others, the authors of the work proposed to take a mean of the results of all methods in order to minimize the risk of large errors. Due to the fact that our classifiers use all features simultaneously as input, in some sense we followed their advice. From our findings it follows that with the full feature set, the machine learning methods applied to the real data yield results completely different from the ones produced with the statistical methods. However, adjusting the diffusion coefficient in the synthetic trajectories to the most frequent value among the real samples or removing this coefficient from the feature set and re-training the classifiers starts to produce results very similar to the $p$-variation method from Ref.~\cite{WER19}.

From the above methods, the one with the reduced set of features is more convenient, because the classifier is trained only once and then may be simply applied to any unseen data. It does not require a recurring and time consuming procedure consisting of:  
(1) adjustment of $\sigma$, (2) generation of tailor-made training data, (3) extraction of features and (4) training of the classifier every time a new set of experimental trajectories is arriving for analysis.

The agreement between the ML approach and   the statistical testing based on $p$ variations is, on one hand, a confirmation that our ML methods are able to classify unseen data in a reasonable way. On the other hand, it may support the choice of the $p$-variation testing procedure among the statistical methods.

Introduction of noise mimicking different kinds of localization errors changed the classification results -- the count of normal diffusion (subdiffusion) trajectories increased (decreased) by a couple of percentage points. A slight increase of superdiffusion samples was also observed in case of G proteins. If we would like to relate the “noisy” results with the ones from Ref.~\cite{WER19}, they are close to an average of $1$-var and $MAX$ methods.

Although still a lot needs to be done in terms of selection of robust features and the generation of appropriate synthetic training data, we believe that our methodology may be successfully applied to experimental data in order to provide a further insight into the dynamics of complex biological processes.

\begin{acknowledgments}
The work of P.K, H. L.-O. J.S. and A.W.  was supported by NCN-DFG Beethoven Grant No. 2016/23/G/ST1/04083. The work of J.J. was supported by NCN Sonata Bis Grant No. 2019/34/E/ST1/00360.
Calculations have been carried out using resources provided by Wroclaw Centre for Networking and Supercomputing (http://wcss.pl).
\end{acknowledgments}

\appendix*
\section{Codes}
Python codes for every stage of the classification procedure shown in Fig.~\ref{fig:workflow}, together with a short documentation, are publicly available at Zenodo (see Ref.~\cite{JAN20}).

\bibliography{spt_ml}

\end{document}